\documentclass[12pt]{article}

\usepackage{graphicx}
\usepackage[sort&compress,comma]{natbib}
\usepackage{fancyhdr}

\renewcommand{\baselinestretch}{1.5}

\makeatletter 
\renewcommand\@biblabel[1]{#1.}
\makeatother

\title{ Systematic Study of the Boundary Composition in               
        Poisson Boltzmann Calculations                    }

\author{ Parimal Kar$^{1}$,
         \and
         Yanjie Wei$^{1}$,
         \and
         Ulrich H. E. Hansmann$^{1,2}$       
         \and
         Siegfried H\"ofinger$^{1*}$ 
       }
\date{ }

\begin{document}
\maketitle
\noindent  
             $^{1}$Department of Physics, Michigan Technological
               University, 1400 Townsend Drive, Houghton, MI,
               49331 - 1295, USA                             \newline
             $^{2}$John v. Neumann Institute for Computing, 
               Forschungszentrum J\"ulich, 52425 J\"ulich, Germany \newline

\vspace{0.5cm}
\noindent  
             $^{*}$Corresponding Author:   

\noindent    
             Siegfried H\"ofinger                            \newline
             Michigan Technological University               \newline
             Department of Physics                           \newline
             1400 Townsend Drive, Houghton, Michigan, 49931 - 1295, USA\newline
             Phone: 1-906-487-1496                           \newline
             Fax:   1-906-487-2933                           \newline
             Email: shoefing@mtu.edu                                  


\thispagestyle{fancyplain}
\lfoot{\vspace*{-2.0cm}  
       \footnotesize
       $\underline{\mbox{\hspace{1.5cm}}}$ \newline
       {\bf Abbreviations:} DNA, desoxyribosenucleic acid; RNA, ribonucleic 
       acid; PB, Poisson Boltzmann; GB, Generalized Born; FDPB, finite 
       difference Poisson Boltzmann; PB/BEM, boundary element method Poisson
       Boltzmann; SASA, solvent accessible surface area; PCM, polarizable
       continuum model; PDB, protein data bank;  PDB-REPRDB, representative 
       protein chains from PDB; MSROLL, molecular surface program by M. 
       Connolly; SIMS, molecular surface program by Y. Vorobjev; MSMS, 
       molecular surface program by M. Sanner; MOLDEN, a molecular graphics
       program; AMBER, a molecular modelling package for biomolecular 
       simulation; DFT, density functional theory; GAUSSIAN-03, a quantum
       chemical software package; GG,AA,VV, di-glycine, di-alanine, di-valine;}
\cfoot{}

\pagebreak

\begin{abstract}
\noindent
We describe a three-stage procedure to analyze the dependence of Poisson
Boltzmann calculations on the shape, size and geometry of the 
boundary between solute and solvent. our study is carried out within the
boundary element formalism, but our results 
are also of interest to finite difference techniques of
Poisson Boltzmann calculations. At first, we identify the critical size 
of the geometrical elements for discretizing the boundary, and thus
the necessary resolution required to establish numerical convergence.
In the following two steps we perform reference calculations on
a set of dipeptides in different conformations using the Polarizable
Continuum Model and a high-level Density Functional as well as a 
high-quality basis set. Afterwards, we propose a mechanism
for defining appropriate boundary geometries. Finally, we compare
the classic Poisson Boltzmann description with the Quantum Chemical
description, and aim at finding appropriate fitting parameters to get 
a close match to the reference data. Surprisingly, 
when using default AMBER partial charges and the rigorous geometric 
parameters derived in the initial two stages, no scaling of the partial
charges is necessary and the best fit against the reference set is 
obtained automatically.
\end{abstract}
\pagebreak

\section{ Introduction }
A common way of describing solvation effects to biomolecular structure is
to treat the solvent as a continuum of characteristic dielectric constant.
The biomolecule of interest, i.e. a protein, DNA, RNA, glycolipid, etc. 
is considered in full atomic detail, while the surrounding medium is 
represented as structureless continuum interacting primarily via 
polarization, dispersion, repulsion and cavitation effects 
\cite{tomasi,cramer,tomasi2,roux,rivail}. The underlying physics concerned 
with polarization is then often expressed in terms of solutions to the 
Poisson-Boltzmann equation (PB) 
\cite{warwicker,honig,gilson,mccammon,karplus,van_gunsteren,zauhar,juffer}. 
Approximations to the PB --- motivated by simplified computational 
protocols --- are  standard practice e.g.  the Generalized Born model (GB) 
\cite{still,case}. However, PB and GB are dealing with the polarization term 
only, and  the other above mentioned interactions are usually treated by 
either first-principle \cite{curutchet} or semi-empirical \cite{jorgensen} 
character.

Solutions to the PB are computed either by the finite difference method 
(FDPB) \cite{warwicker,honig,gilson,mccammon} or by the boundary element 
method (PB/BEM) \cite{zauhar,juffer}. The latter is particularly intriguing 
since it reduces a three-dimensional integral over the entire volume to a 
two-dimensional surface integral, leading to considerable savings in 
computational time. Both approaches depend  fundamentally on the exact  
definition of the boundary between solute and solvent. All definitions are 
based on the area of the atoms exposed to the solvent, for instance the 
solvent accessible surface area (SASA), the solvent excluded volume, or the 
molecular surface \cite{connolly}, which all depend on a chosen set 
of van der Waals radii \cite{lii,aqvist,weiner,halgren} assigned to the 
center of the atoms.

Given the dependence on the exact geometry and quality of the boundary it 
appears necessary to study the geometric factors that influence the outcome 
of PB calculations in greater detail. This is particularly appropriate for 
semi-quantitative approaches \cite{kollman} where the demand on accuracy is 
a very sensitive issue \cite{bates}. Particular attention has to be drawn to 
factors such as
i)   surface type and surface resolution,       
ii)  dependence on atomic model parameters, i.e. van der Waals radii,
iii) generality and physicochemical significance. In this present work 
we provide such an analysis by focussing on each of these three points 
separately. At first, we employ different surface generation algorithms to 
a subset of randomly chosen protein structures of variable size and shape. 
PB/BEM calculations are carried out with increasing resolution of the 
boundary. Optimal surface resolution and surface generation parameters that 
guarantee numerical convergence and methodic stability are derived.
Next, we use these optimized parameters for a set of model peptides and vary 
the van der Waals radii in a systematic way. The reference set of model 
peptides is considered at a high level of quantum chemical theory, i.e. 
PCM \cite{tomasi2} using the Becke-98 density functional \cite{becke} and the 
basis set of Sadlej \cite{sadlej}. The aim of this second step is to identify 
optimal van der Waals radii within the PB/BEM approach that will lead to 
boundaries and solute geometries of similar size and shape as those used in 
the high-level PCM calculations. Finally, with the optimized parameters 
determined in the initial two stages we  compute actual PB/BEM polarization 
energies in order to obtain a close match with the quantum chemical results 
obtained from the reference set.

\section{ Methods }
\subsection{ Sample Selection, Preparation and Set Up of Structures and 
             Computation of Molecular Surfaces with Different Programs   } 
A set of different protein structures is randomly selected from the 
Protein Data Bank \cite{pdb}. The actual download site used is the
repository PDB-REPRDB \cite{pdb1}. Default options are applied with the 
following exceptions: 
i)   {\bf Number of residues less than 40 excluded -- NO},
ii)  {\bf Include MUTANT -- NO},
iii) {\bf Exclude COMPLEX},
iv)  {\bf Exclude FRAGMENT},
v)   {\bf Include NMR -- NO},
vi)  {\bf Include Membrane Proteins -- NO}.
A total of 28 structures 
of different protein sizes and shapes (see table \ref{table1}) are chosen. 
The PDB codes of the samples are,
2ERL, 1P9GA, 1FD3A, 1N13E, 1BRF, 1PARB, 1K6U, 1AVOA, 1SCMA, 1OTFA, 1DJTA,  
1KU5, 1K3BC, 1R2M, 1CC8, 1L9LA, 1ZXTD, 1GYJA, 1T8K, 1XMK, 1YNRB, 1EZGA,  
1C5E, 1SAU, 1WN2, 1JBE, 1C7K and 1WKR.

Four different programs to calculate molecular surfaces have been employed:
the Connolly program {\bf MSROLL} \cite{connolly}, the {\bf SIMS} program 
\cite{vorobjev}, the {\bf MSMS} program by Sanner \cite{sanner} and a 
self-written program based on an estimation of the SASA \cite{skrivanek}. 
However, we detected already in early stages of this investigations  
problems with varying van der Waals radii and found indications that the 
SASA is not an appropriate choice for this kind of application. Hence, we 
have dropped the latter two programs, and most of the analysis is done with 
programs {\bf MSROLL} and {\bf SIMS}. 

Downloaded PDB structures are cleaned from multichain entries, HETATM 
lines CONNECT lines, ANISOU lines, counter ions, water molecules and the 
footer section. Program MOLDEN \cite{schaftenaar} is used to visualize the 
downloaded PDB structures after cleaning and the force field {\bf Tinker Amber}
is selected before a new PDB file is written out from within MOLDEN using 
option 'Write\_With\_Hydrogens'. Since MOLDEN always uses the default 
HIP-type in AMBER jargon, HIS residues need to be converted to HIP types, as 
well as CYS residues engaged in disulfide bonds need to be converted to 
CYX-type residues. Occasional cases with PRO being the initial residue are 
manually edited and initial PROs removed. AMBER non-bonded parameters 
\cite{cornell}, i.e. charges and van der Waals radii are assigned to all the 
atoms in the protein structures. In this first part of the study, the 
vdW-radii are increased by a factor of 1.12 and atomic partial charges are 
scaled down by another factor of 0.9 \cite{hoefi}.

The MSROLL program is used with varying choices of the fineness value
(the {\bf -f} command line argument) which  defines the resolution of the 
surface. With smaller values the resolution of the surface becomes better 
but computational cost will increase. The probe radius (the {\bf -p} command 
line argument) is set to 1.5 \AA.  Analytically calculated SASA and molecular 
volumes are recorded, and the data file containing triangulation details is 
translated into a human readable format, and critical items (for example
almost coinciding triangles) removed.

The SIMS program is used with identical arguments to those employed in
MSROLL. Similarly, varying the resolution of the surface triangulation
into small sized triangles means adjusting the {\bf dot-density} parameter 
in SIMS. Higher values for this parameter will yield higher surface 
resolutions but also increase the computational demand. We record the 
number of BE, number of iterations, SASA and volume for comparison.

\subsection{ Computation of Polarization Free Energies, $\Delta G^{Pol}$,
             Based on solutions to the Poisson Boltzmann Equation        }
Inner/outer dielectric constants at the molecular boundary are set to 1.0 
and 80.0 respectively. The serial version of the PB/BEM program {\bf POLCH} 
\cite{hoefi1} is used. Critical cases with additional secondary cavities 
located in the interior part of the proteins are excluded. AMBER van der 
Waals radii and partial charges \cite{cornell} are applied. Using our own 
tool chain for the assignment allows us to conveniently scale these data, 
as well as to write out in the same instance the corresponding parameter 
files required by the molecular surface programs.

\subsection{ Density Functional Theory Calculations Including the 
             Polarizable Continuum Model on a Reference Set of 
             Dipeptides                                                  }
The most prominent combinations of peptidic $\Phi,\Psi$-angles \cite{marshall} 
are used to construct different conformations of dipeptides. Only homodimers
are considered. All 20 types of different amino acids are used for this 
combinatorial approach. Zwitter-ionic forms are built and 9 conformations per 
class of amino acid are taken into account leading to all in all 180 
structures. Program ``protein.x'' from the TINKER package version 4.2 is
employed \cite{tinker}. Each of these reference structures is subjected to PCM 
\cite{tomasi2} calculations at the Becke-98 \cite{becke} level of density 
functional theory (DFT) using the high-quality basis set of Sadlej 
\cite{sadlej} within the Gaussian-03 suite of programs \cite{gauss03}.
Geometric properties, i.e. the molecular volume and the molecular surface 
area, as well as polarization free energies are extracted from the reference
calculations and used as a base line when comparing to PB/BEM data. 
The computational demand of these reference calculations is significant.
For example, WW-conformations require on the order of 6 weeks (and beyond)
single-processor time on modern computing architectures.

\section{ Results }
\subsection{ Stage I: Rather small-sized BEs are Needed to Obtain Consistently
             Convergent Polarization Free Energies $\Delta G^{Pol}$ }
\begin{figure}[ht]
\begin{center}
\includegraphics[scale=1.0]{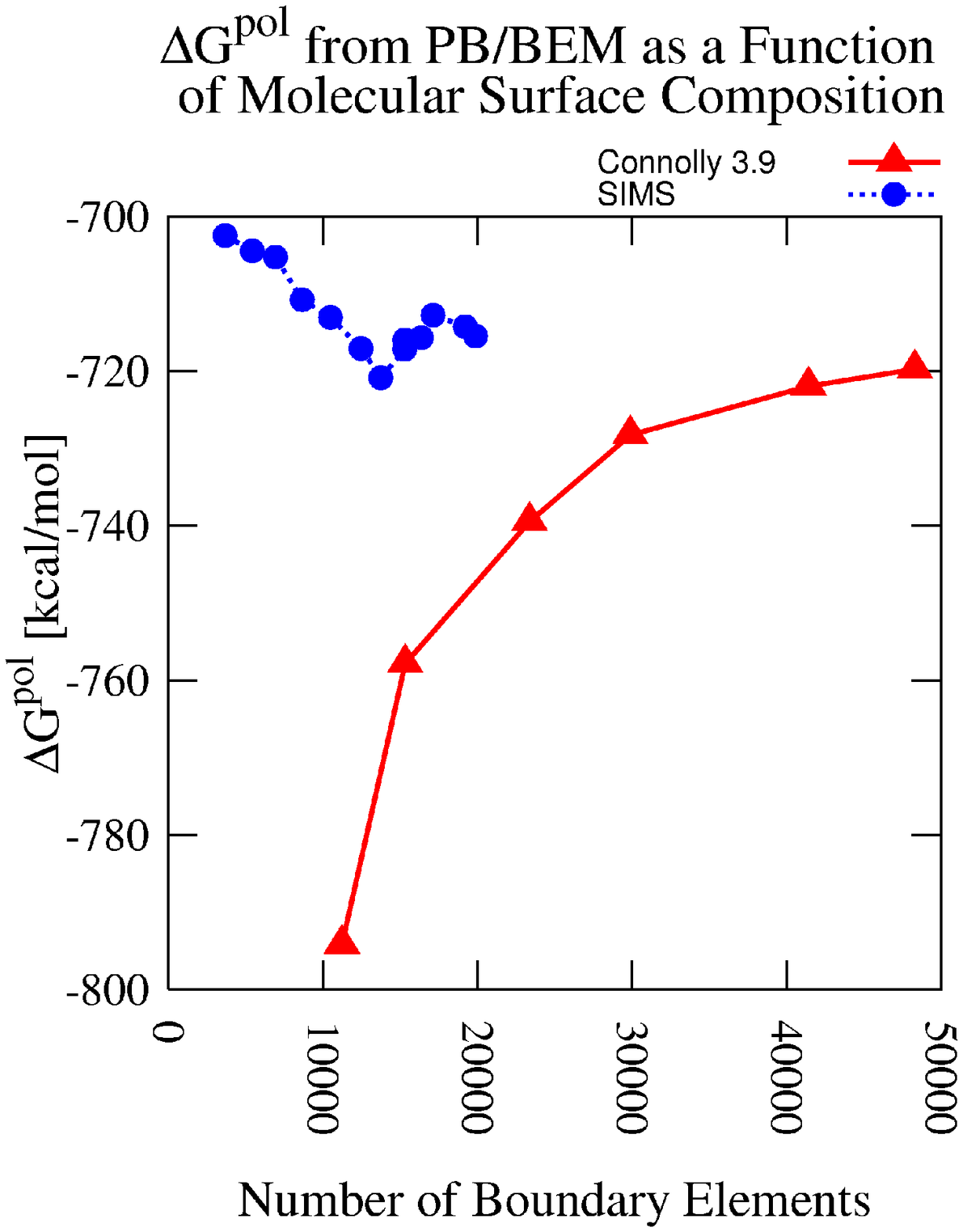}
\caption[  PB/BEM derived polarization free energies $\Delta G^{Pol}$ as
           a function of boundary resolution obtained from two independent
           programs MSROLL \cite{connolly} and SIMS \cite{vorobjev}. The
           example represents results for PDB structure 1C5E.            ]
        { \label{figure1}                                                  }
\end{center}
\end{figure}

We start with PB/BEMA calculations for a set of protein structures 
(PDB codes summarized in Table \ref{table1}).  The boundary discretization is 
achieved with two independent programs, MSROLL \cite{connolly} and SIMS 
\cite{vorobjev}. Boundary resolution into BEs is steadily increased with 
either program and independent PB/BEM results are computed for each particular 
boundary decomposition. A typical plot of the trend of $\Delta G^{Pol}$ as
a function of number of BEs is shown in Figure \ref{figure1} for the protein 
structure with PDB code 1C5E. Similar plots for the other examples in Table 
\ref{table1} are provided as supplementary material. Both approaches converge 
to identical results in the limit of large numbers of BEs. The importance of 
well-resolved boundaries becomes clear  from  Figure \ref{figure1}. 
Errors on the order of $\pm 40 \frac{kcal}{mol}$ are easily introduced
when working in the non-converged domain. Connolly's MSROLL program (red 
triangles in Figure \ref{figure1}) reaches a plateau value in a continuous 
manner, while the SIMS program (blue spheres in Figure \ref{figure1}) finds 
its limit value within an alternating sequence. The SIMS program reaches 
convergence much faster than the Connolly program. The quality of the computed 
molecular boundaries is comparable, see, for instance, the values of molecular 
surfaces and volumes (final two columns in Table \ref{table1}) obtained with 
either program. SIMS seems to overestimate the volume by a small margin of 
roughly $1 \%$.
The recommended average size of BEs for converged results using MSROLL
is on the order of 0.11 \AA$^2$ while SIMS would require an average size
of 0.31 \AA$^2$. Both numbers are close to the value of 0.4 \AA$^2$ advocated 
in Quantum Chemistry \cite{esqc-00}.

\subsection{ Stage II: Systematic Geometric Comparison to High Level Quantum 
             Chemistry Calculations Suggests a Uniform Scaling of AMBER van 
             der Waals Radii by a Factor of 1.07                             }
\begin{figure}[ht]
\begin{center}
\includegraphics[scale=1.0]{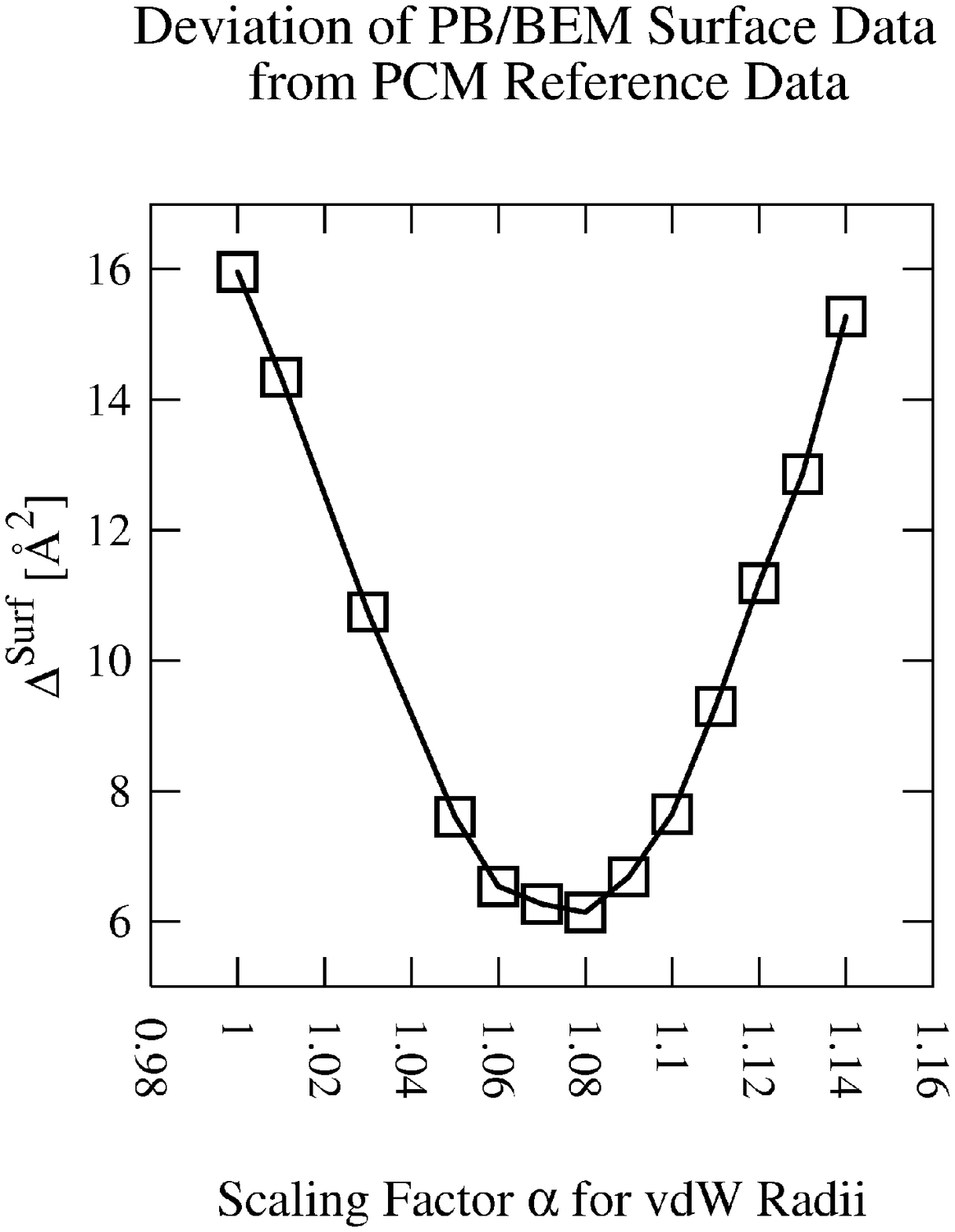}
\caption[  Comparison of employed molecular surfaces in the PB/BEM series
           based on scaling the AMBER default van der Waals radii by a 
           factor $\alpha$ to the reference data obtained from PCM 
           calculations at the Becke-98/Sadlej level.                    ]
        { \label{figure2}                                                  }
\end{center}
\end{figure}

\begin{figure}[ht]
\begin{center}
\includegraphics[scale=1.0]{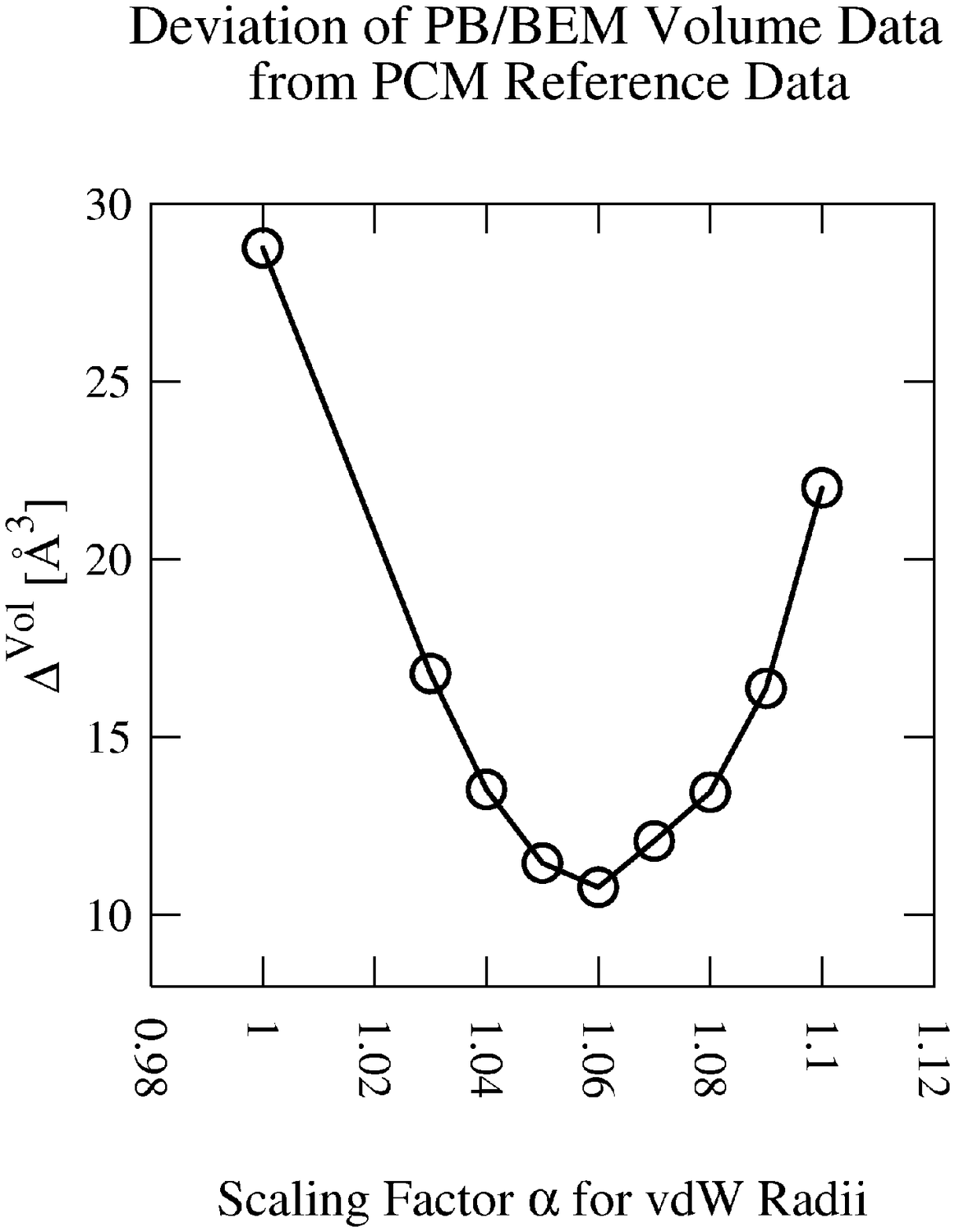}
\caption[  Comparison of employed molecular volumes in the PB/BEM series
           based on scaling the AMBER default van der Waals radii by a 
           factor $\alpha$ to the reference data obtained from PCM 
           calculations at the Becke-98/Sadlej level.                    ]
        { \label{figure3}                                                  }
\end{center}
\end{figure}

A reference set of dipeptides in different conformations (9 per species) is 
constructed. Only homodipeptides comprising all 20 types of naturally 
occurring amino acids are considered. Thus a total number of 180 dipeptidic
reference structures is set up. The zwitterionic form is used throughout.
Each of these structures is computed at the Becke-98 level of theory 
\cite{becke} using the basis set of Sadlej \cite{sadlej} and the PCM model
\cite{tomasi2} for solvation free energies. Geometric properties such as
the cavity volume and the cavity surface area are extracted from each of 
the reference calculations. All 180 structures are also computed within the
PB/BEM approach using optimized parameters for the boundary resolution 
determined in Stage I of this study. However, only the SIMS program is used. 
We define a global deviation  from the reference 
data by
\begin{equation}
\label{eq1}
  \Delta^{Surf} = \frac{1}{20}
                  \sum\limits_{i=1}^{20}
                  \frac{1}{9}
                  \sum\limits_{j=1}^{9}
                  \sqrt{ (Surf_{i,j}^{PCM} - Surf_{i,j,\alpha}^{PB/BEM})^2 }
\end{equation}
where $j$ runs over the conformations and $i$ over the different types of 
homodipeptides, i.e. GG, AA, VV, etc. The parameter  $\alpha$ refers to a 
specific scaling factor used when constructing the boundaries within the 
PB/BEM approach. In particular this scaling makes the  van der Waals radii
larger or smaller by a certain fraction. The AMBER default set of van der
Waals radii is used \cite{cornell}. A similar criterion is used for comparing 
molecular volumes,
\begin{equation}
\label{eq2}
  \Delta^{Vol} = \frac{1}{20}
                  \sum\limits_{i=1}^{20}
                  \frac{1}{9}
                  \sum\limits_{j=1}^{9}
                  \sqrt{ (Vol_{i,j}^{PCM} - Vol_{i,j,\alpha}^{PB/BEM})^2 }
\end{equation}
and the dependence on the scaling factor $\alpha$ is shown in Figures
\ref{figure2} and \ref{figure3}. As becomes clear from Figures \ref{figure2} 
and \ref{figure3} the best match to the reference data is obtained when 
scaling the AMBER van der Waals radii by a factor of 1.07. Detailed data
with respect to conformational averages per type of dipeptide 
are shown in Tables \ref{table2} and \ref{table3}.

\subsection{ Stage III: Charge Scaling is Not Required }                      
\begin{figure}[ht]
\begin{center}
\includegraphics[scale=1.0]{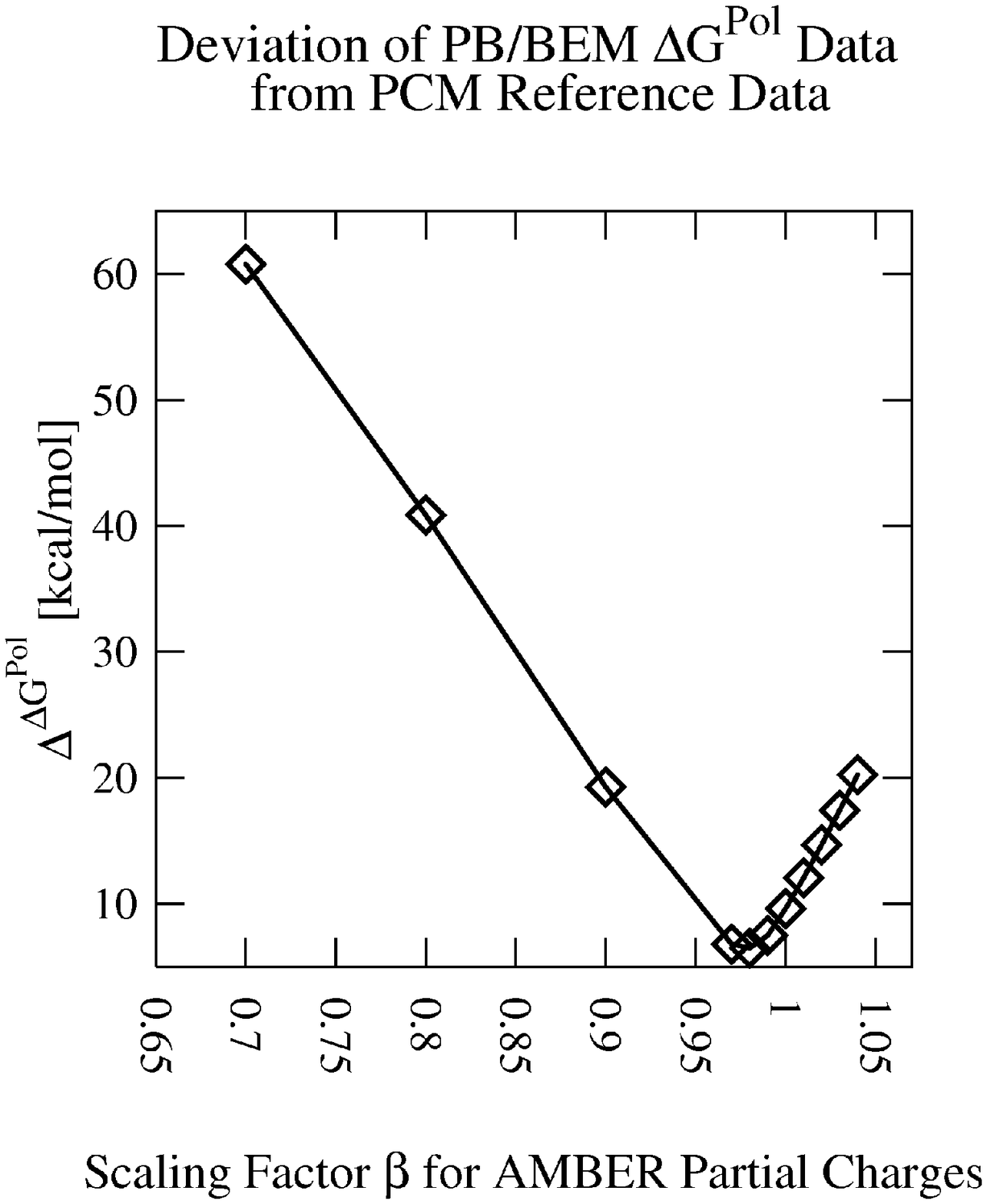}  
\caption[  Comparison of PB/BEM polarization free energies $\Delta G^{Pol}$
           based on scaling the AMBER default charges by a factor $\beta$  
           to the reference data obtained from PCM calculations at the 
           Becke-98/Sadlej level.                                         ]
        { \label{figure4}                                                  }
\end{center}
\end{figure}

Using  the optimized parameters obtained in the previous two stages leads us 
to the final step of directly comparing polarization free energies 
$\Delta G^{Pol}$ computed within the PB/BEM approximation and at the PCM level 
of theory. The idea is to identify another uniform scaling factor $\beta$ 
which applied to the AMBER default charges would result in an optimal match 
to the reference polarization free energies. Thus another deviation criterion 
is introduced,                
\begin{equation}
\label{eq3}
  \Delta^{\Delta G^{Pol}} = \frac{1}{20}
                  \sum\limits_{i=1}^{20} 
                  \frac{1}{9} 
                  \sum\limits_{j=1}^{9}
                  \sqrt{ (\Delta G_{i,j}^{Pol,PCM} 
                          - \Delta G_{i,j,\beta}^{Pol,PB/BEM})^2 }
\end{equation}
that allows to identify  the optimal value of $\beta$. The dependence of
the PB/BEM polarization free energies on the charge scaling factor 
$\beta$ is shown in Figure \ref{figure4}. The trend shown in Figure 
\ref{figure4} suggests an optimal value of $\beta$ very close to 1.0,
hence no charge scaling is required. This result  i) emphasizes the broad 
applicability of AMBER partial charges and ii) circumvents conceptual 
difficulties that would arise when charges had to be scaled, i.e. 
modified net charges in proteins, non-neutral forms, etc. A detailed
analysis with respect to the magnitude of the average deviation of 
each particular type of dipeptide studied is shown in Table \ref{table4}.

\section{ Discussion }
Motivated by our recent high-performance solution to Poisson Boltzmann 
calculations \cite{hoefi1} we have now tested the influence of the many 
critical parameters involved. One obvious issue is the exact choice and 
composition of the boundary between solute and solvent. At first, we have to 
ensure the numerical stability within the selected level of approximation.
In order to address this problem we have carried out PB/BEM calculations on 
a large sample of different proteins. When using different programs to create 
the boundary surface and increasing systematically the resolution of these 
surfaces into small-sized boundary elements, a recommended threshold size of 
about 0.31 \AA$^2$ for the average BE is identified when using program 
SIMS \cite{vorobjev} which showed faster convergence than the well-known 
Connolly program \cite{connolly}. Although giving rise to very fine-resolved 
boundary surfaces, hence large numbers of BEs, this value is close to the 
corresponding value of 0.4 \AA$^2$ frequently advised in Quantum Chemical 
models \cite{esqc-00}. As a consequence, even proteins of modest size thus 
require consideration of vast numbers of BEs (see for example Table 
\ref{table1}), and the importance of efficient means of solving the 
computational problem is underlined again. 

After having established the necessary degree of boundary partitioning
in the first stage, we performed a systematic comparison against a reference
set of dipeptides computed at a high level of Quantum Chemical theory.
Consideration of geometric factors revealed that when applying a scaling 
factor of about 1.07 to AMBER default van der Waals radii, rather good  
agreement can be reached between the reference geometries and the geometries 
in the PB/BEM approach. The recommended value of 1.07 is somewhat smaller than 
a factor found previously (1.12 of ref \cite{hoefi}) and reflects the much 
finer resolved boundary surfaces used in this present work.

The final step was to compare actual calculations of the polarization free 
energies to each other. Following previous attempts, we wanted to derive 
another scaling factor that, when applied to AMBER partial charges, would 
yield a close match to the reference polarization free energies. The trend 
visible  in Figure \ref{figure4} indicates that no scaling of the charges is 
necessary: they are already close to optimal. This is an unexpected 
--- but very welcome --- result, as it eliminates potential secondary    
problems that would emerge with modifying charges. Again, this is another
consequence of the much finer resolved boundary surfaces in this present 
work as opposed to previous results \cite{hoefi} where a scaling
factor of 0.9 had been found.

\section{ Conclusion }
Combined employment of small-sized BEs ($\approx$ 0.3 \AA$^2$ on average), 
slightly increased AMBER van der Waals radii (by a factor of 1.07), and default 
AMBER partial charges leads to good quality estimates of the polarization 
free energy, $\Delta G^{Pol}$, for proteins within the PB/BEM framework.

{\bf\large Acknowledgements }
This work was supported in part by the National Institutes of Health 
Grant GM62838 and by a grant of the National Science Foundation (CHE-0313618).
Dr. Michael Connolly is acknowledged for providing a test version of his 
molecular surface program.
\pagebreak


\clearpage
\listoffigures

\clearpage
\thispagestyle{empty} 
\clearpage

\clearpage
\thispagestyle{empty}
\clearpage

\clearpage
\thispagestyle{empty}
\clearpage

\clearpage
\thispagestyle{empty}
\clearpage

\clearpage
\thispagestyle{empty}
\renewcommand{\baselinestretch}{1.3}
\begin{table}
\begin{center}
\vspace*{-2.0cm}
\caption{\label{table1} 
         PDB codes of studied structures and corresponding number of 
         residues. Columns three and four show the number of BEs needed to 
         reach converged PB/BEM results using molecular surface algorithms
         MSROLL and SIMS respectively. Final two columns show average
         molecular surface areas and average molecular volumes derived when
         using data of both programs, MSROLL and SIMS. The differences 
         obtained when subtracting the SIMS results from the MSROLL results 
         are given in parentheses.                                            }
\begin{tabular}{lrrrrr}
\hline
  \multicolumn{1}{c}{ PDB Code }                     &
  \multicolumn{1}{c}{ Number of }                    &
  \multicolumn{1}{c}{ Number of BEs }                &
  \multicolumn{1}{c}{ Number of BEs }                &
  \multicolumn{1}{c}{ Molecular }                    &
  \multicolumn{1}{c}{ Solvent  }                     \\
                                                     &
  \multicolumn{1}{c}{ Residues }                     &
  \multicolumn{1}{c}{ at Convergence }               &
  \multicolumn{1}{c}{ at Convergence }               &
  \multicolumn{1}{c}{ Surface Area }                 &
  \multicolumn{1}{c}{ Excluded Volume }              \\
                                                     &
                                                     &
  \multicolumn{1}{c}{ Using MSROLL }                 &
  \multicolumn{1}{c}{ Using SIMS }                   &
  \multicolumn{1}{c}{ (Difference) }                 &
  \multicolumn{1}{c}{ (Difference) }                 \\
                                                     &
                                                     &
  \multicolumn{1}{c}{ \cite{connolly} }              &
  \multicolumn{1}{c}{ \cite{vorobjev} }              &
  \multicolumn{1}{c}{ [\AA$^2$] }                    &
  \multicolumn{1}{c}{ [\AA$^3$] }                    \\
\hline
  2ERL   &   40    &   15661   &   9807  &  2370 (+1)   &  5653 (-43)   \\
  1P9GA  &   41    &   22302   &   5751  &  2091 (-5)   &  5055 (-72)   \\
  1FD3A  &   44    &   25865   &   6699  &  2408 (+7)   &  5819 (-56)   \\
  1N13E  &   52    &   18419   &  10353  &  3750 (+11)  &  6542 (-69)   \\
  1BRF   &   53    &   33879   &  11810  &  2796 (-7)   &  7734 (-77)   \\
  1PARB  &   53    &   42336   &  11006  &  3968 (-8)   &  8509 (-108)  \\
  1K6U   &   58    &   24220   &  13406  &  3195 (+12)  &  8603 (-65)   \\
  1AVOA  &   60    &   43916   &  13335  &  4777 (-3)   &  9325 (-186)  \\
  1SCMA  &   60    &   54464   &  14603  &  5131 (-13)  & 10601 (-179)  \\
  1OTFA  &   62    &   40128   &  10610  &  3767 (+6)   &  8942 (-86)   \\
  1DJTA  &   64    &   35828   &   9134  &  3331 (+26)  &  9422 (-43)   \\
  1KU5   &   66    &   46390   &  12208  &  4310 (+3)   & 10153 (-133)  \\
  1K3BC  &   69    &   61667   &  16297  &  5768 (+19)  & 18193 (-165)  \\
  1R2M   &   71    &   39316   &  13659  &  3244 (-1)   &  9596 (-74)   \\
  1CC8   &   73    &   27668   &  15091  &  3644 (+2)   & 11094 (-65)   \\
  1L9LA  &   74    &   20278   &  11636  &  4182 (+5)   & 11728 (-112)  \\
  1ZXTD  &   76    &   37335   &  11259  &  4089 (+8)   & 10809 (-93)   \\
  1GYJA  &   76    &   44770   &  13665  &  4885 (-3)   & 11464 (-118)  \\
  1T8K   &   77    &   35978   &  13846  &  3925 (+4)   & 11410 (-119)  \\
  1XMK   &   79    &   56033   &  16468  &  4294 (+9)   & 12288 (-98)   \\
  1YNRB  &   80    &   31529   &  12630  &  4417 (+16)  & 11911 (-157)  \\
  1EZGA  &   84    &   34628   &   9122  &  3258 (+3)   & 10103 (-95)   \\
  1C5E   &   95    &   48306   &  19880  &  4480 (0)    & 13285 (-110)  \\
  1SAU   &  115    &   47613   &  17765  &  5197 (+25)  & 17897 (-116)  \\
  1WN2   &  121    &   51325   &  21555  &  5614 (+15)  & 17836 (-118)  \\
  1JBE   &  128    &   58119   &  16729  &  5409 (+22)  & 18905 (-188)  \\
  1C7K   &  132    &   54104   &  16675  &  5389 (0)    & 18858 (-182)  \\
  1WKR   &  340    &   74167   &  55378  & 11008 (-41)  & 47105 (-299)  \\
\hline
\end{tabular}
\end{center}
\end{table}
\clearpage

\clearpage
\thispagestyle{empty}
\renewcommand{\baselinestretch}{1.5}
\begin{table}
\begin{center}
\caption{\label{table2} 
         Comparison of average molecular surfaces based on unscaled 
         and scaled AMBER van der Waals radii with data from PCM
         reference calculations. The average comprises 9 different
         conformations per type of dipeptide considered. The value in 
         parentheses monitors that conformation that deviates most
         severely from the mean.                                      }
\begin{tabular}{lrrr}
\hline
  \multicolumn{1}{c}{ Dipeptide }                    &
  \multicolumn{1}{c}{ Mean Surface }                 &
  \multicolumn{1}{c}{ Mean Surface }                 &
  \multicolumn{1}{c}{ Mean Surface }                 \\
  \multicolumn{1}{c}{ Type }                         &
  \multicolumn{1}{c}{ AMBER Unscaled }               &
  \multicolumn{1}{c}{ AMBER Scaled }                 &
  \multicolumn{1}{c}{ PCM Reference }                \\
                                                     &
  \multicolumn{1}{c}{ [\AA$^2$] }                    &
  \multicolumn{1}{c}{ [\AA$^2$] }                    &
  \multicolumn{1}{c}{ [\AA$^2$] }                    \\
\hline
 AA  &    191.764 (5.205)   &      204.388 (3.697)   &    214.747 (4.955)  \\
 CC  &    214.592 (4.622)   &      229.274 (5.461)   &    232.910 (6.522)  \\
 DD  &    228.687 (6.032)   &      242.724 (5.972)   &    240.839 (6.776)  \\
 EE  &    273.402 (5.835)   &      287.557 (6.951)   &    283.025 (5.766)  \\
 GG  &    150.255 (4.460)   &      161.637 (4.936)   &    167.764 (3.249)  \\
 II  &    279.535 (10.274)  &      294.402 (11.331)  &    302.173 (12.233) \\  
 KK  &   314.712 (6.273)    &      332.593 (7.677)   &    340.545 (7.599)  \\
 LL  &   276.435 (10.012)   &      290.497 (12.440)  &    293.172 (10.465) \\
 MM  &   301.384 (6.691)    &      318.033 (8.261)   &    329.815 (8.374)  \\
 NN  &   232.466 (6.072)    &      247.553 (7.325)   &    247.040 (7.310)  \\
 QQ  &   276.939 (5.957)    &      293.663 (7.531)   &    292.648 (6.582)  \\
 RR  &   354.636 (6.925)    &      377.310 (7.568)   &    380.408 (6.739)  \\
 SS  &   196.528 (4.982)    &      207.904 (4.610)   &    212.107 (4.695)  \\
 TT  &   224.181 (8.047)    &      238.570 (8.359)   &    239.112 (10.124) \\
 VV  &   251.913 (8.574)    &      265.008 (8.296)   &    276.423 (8.140)  \\
 YY  &   340.272 (17.100)   &      356.042 (17.293)  &    346.378 (14.704) \\
 FF  &   329.058 (17.123)   &      343.245 (17.402)  &    326.947 (15.489) \\
 WW  &   355.790 (26.425)   &      377.209 (27.865)  &    361.573 (25.182) \\
 HH  &   282.802 (13.007)   &      299.235 (12.829)  &    296.885 (11.461) \\
 PP  &   224.999 (10.768)   &      237.525 (10.500)  &    233.118 (9.900)  \\
\hline
\end{tabular}
\end{center}
\end{table}
\clearpage

\clearpage
\thispagestyle{empty}
\begin{table}
\begin{center}
\caption{\label{table3} 
         Comparison of average molecular volumes based on unscaled 
         and scaled AMBER van der Waals radii with data from PCM
         reference calculations. The average comprises 9 different
         conformations per type of dipeptide considered. The value in 
         parentheses monitors that conformation that deviates most
         severely from the mean.                                      }
\begin{tabular}{lrrr}
\hline
  \multicolumn{1}{c}{ Dipeptide }                    &
  \multicolumn{1}{c}{ Mean Volume }                  &
  \multicolumn{1}{c}{ Mean Volume }                  &
  \multicolumn{1}{c}{ Mean Volume }                  \\
  \multicolumn{1}{c}{ Type }                         &
  \multicolumn{1}{c}{ AMBER Unscaled }               &
  \multicolumn{1}{c}{ AMBER Scaled }                 &
  \multicolumn{1}{c}{ PCM Reference }                \\
                                                     &
  \multicolumn{1}{c}{ [\AA$^3$] }                    &
  \multicolumn{1}{c}{ [\AA$^3$] }                    &
  \multicolumn{1}{c}{ [\AA$^3$] }                    \\
\hline
 AA  & 191.804 (4.553)   &   215.661 (3.355)   &    228.591 (2.885)  \\
 CC  & 223.713 (3.799)   &   252.135 (4.069)   &    255.902 (3.594)  \\
 DD  & 242.406 (4.242)   &   270.928 (5.037)   &    260.904 (4.723)  \\
 EE  & 296.037 (4.888)   &   327.303 (3.864)   &    311.645 (3.810)  \\
 FF  & 381.346 (5.875)   &   417.696 (7.992)   &    388.363 (6.223)  \\
 GG  & 136.315 (3.565)   &   154.314 (2.631)   &    164.287 (2.048)  \\
 HH  & 317.309 (3.605)   &   353.465 (5.228)   &    340.591 (5.324)  \\
 II  & 323.590 (6.689)   &   357.080 (7.564)   &    367.182 (6.520)  \\
 KK  & 343.720 (3.747)   &   382.391 (4.384)   &    388.641 (3.903)  \\
 LL  & 313.537 (5.063)   &   346.507 (6.450)   &    344.950 (7.458)  \\
 MM  & 325.073 (5.069)   &   363.563 (5.484)   &    377.789 (3.905)  \\
 NN  & 248.729 (4.628)   &   278.449 (4.968)   &    271.305 (5.415)  \\
 PP  & 242.861 (8.154)   &   269.561 (9.116)   &    263.128 (9.957)  \\
 QQ  & 301.279 (3.985)   &   336.468 (5.590)   &    325.660 (5.591)  \\
 RR  & 384.833 (4.089)   &   431.811 (4.444)   &    424.874 (3.559)  \\
 SS  & 198.981 (3.533)   &   221.590 (3.162)   &    225.193 (2.964)  \\
 TT  & 244.397 (5.673)   &   272.848 (7.051)   &    273.546 (6.533)  \\
 VV  & 282.296 (6.114)   &   311.383 (6.895)   &    330.378 (8.172)  \\
 WW  & 431.248 (14.910)  &   480.974 (17.118)  &    447.693 (15.019) \\
 YY  & 393.622 (5.433)   &   433.845 (7.433)   &    407.695 (5.466)  \\
\hline
\end{tabular}
\end{center}
\end{table}
\clearpage

\clearpage
\thispagestyle{empty}
\begin{table}
\begin{center}
\caption{\label{table4} 
         Comparison of average PB/BEM polarization free energies 
         $\Delta G^{Pol}$ using AMBER default charges to corresponding
         data obtained from PCM reference calculations. The average 
         comprises a variable number of conformations for each different
         type of dipeptide considered depending on how many PCM 
         calculations terminated faithfully. Numbers in parentheses
         reflect maximum deviation from the mean.                       }
\begin{tabular}{lrrrr}
\hline
  \multicolumn{1}{c}{ Dipeptide }                    &
  \multicolumn{1}{c}{ Mean $\Delta G^{Pol,PB/BEM}$ } &
  \multicolumn{1}{c}{ Mean $\Delta G^{Pol,PCM}$ }    &
  \multicolumn{1}{c}{ Mean $\Delta\Delta G^{Pol}$ }  &
  \multicolumn{1}{c}{ Number of }                    \\
  \multicolumn{1}{c}{ Type }                         &
  \multicolumn{1}{c}{ AMBER Default Charges }        &
  \multicolumn{1}{c}{ PCM Reference }                &
  \multicolumn{1}{c}{ Deviation }                    & 
  \multicolumn{1}{c}{ References }                   \\
                                                     &
  \multicolumn{1}{c}{ [kcal/mol] }                   &
  \multicolumn{1}{c}{ [kcal/mol] }                   &
  \multicolumn{1}{c}{ [kcal/mol] }                   &
                                                     \\
\hline
 AA &  -91.36 ( 8.56 ) &  -83.89 (10.12 ) &   7.47 & 9 \\
 CC & -115.11 (11.02 ) &  -96.80 (12.83 ) &  18.31 & 9 \\
 DD & -296.25 (17.08 ) & -285.27 (18.24 ) &  10.98 & 9 \\
 EE & -266.54 (14.09 ) & -259.29 (13.76 ) &   7.25 & 9 \\
 GG &  -96.52 (10.09 ) &  -89.41 (11.36 ) &   7.11 & 9 \\
 II &  -82.72 ( 7.49 ) &  -75.97 ( 8.70 ) &   6.77 & 9 \\
 KK & -249.63 (16.51 ) & -236.37 (19.64 ) &  13.26 & 9 \\
 LL &  -85.54 ( 7.20 ) &  -64.51 ( 8.64 ) &  21.03 & 9 \\
 MM &  -88.82 ( 7.55 ) &  -82.10 ( 9.42 ) &   6.72 & 9 \\
 NN & -105.11 ( 8.19 ) & -101.80 (12.20 ) &   4.25 & 9 \\
 QQ & -119.08 (10.88 ) & -115.33 (12.60 ) &   3.89 & 9 \\
 RR & -235.39 (17.79 ) & -228.71 (21.45 ) &   6.68 & 6 \\
 SS & -112.78 (13.38 ) & -105.47 (13.90 ) &   7.32 & 9 \\
 TT & -106.87 (12.06 ) & -100.61 (12.88 ) &   6.55 & 9 \\
 VV &  -85.17 ( 7.44 ) &  -77.46 ( 8.73 ) &   7.70 & 9 \\
 YY &  -93.35 ( 4.36 ) &  -90.11 ( 8.48 ) &   3.59 & 5 \\
 FF &  -89.92 (10.55 ) &  -82.51 (13.94 ) &   7.41 & 6 \\
 HH & -237.74 (19.08 ) & -236.06 (22.15 ) &   3.66 & 9 \\
 PP &  -79.15 ( 5.73 ) &  -82.71 ( 7.50 ) &   3.56 & 9 \\
 WW & -100.50 ( 4.27 ) &  -88.11 (12.93 ) &  12.39 & 2 \\
\hline
\end{tabular}
\end{center}
\end{table}
\clearpage

\clearpage
\listoffigures


\clearpage

\begin{thebibliography}{99}
\bibitem[1]{tomasi}
Tomasi, J.;  Persico, M.
Chem Rev 1994, 94, 2027. 
%
\bibitem[2]{cramer}
Cramer, C. J.;  Truhlar, D. G.
Chem Rev 1999, 99, 2161. 
%
\bibitem[3]{tomasi2}
Tomasi, J.; Mennucci, B.; Cammi, R.
Chem Rev 2005,  105, 2999.
%
\bibitem[4]{roux}
Roux, B.; Simonson, T.                 
Biophys Chem 1999,  78, 1.
%
\bibitem[5]{rivail}
Rinaldi, D.; Ruiz-L\'{o}pez, M. F., Rivail, J. L.
J Chem Phys 1983, 78, 834.      
%
\bibitem[6]{warwicker}
Warwicker, J.; Watson, H. C. 
J Mol Biol 1982, 157, 671.
%
\bibitem[7]{honig}
Honig, B.; Nicholls, A.    
Science 1995, 268, 1144.                                       
%
\bibitem[8]{gilson}
Luo, R.; David, L.; Gilson, M. K.                              
J Comput Chem 2002, 23, 1244.                       
%
\bibitem[9]{mccammon}
Baker, N. A.; Sept, D.; Joseph; S.; Holst, M. J.; McCammon, J. A. 
Proc Natl Acad Sci USA 2001, 98, 10037.          
%
\bibitem[10]{karplus}
Bashford, D.;  Karplus, M.
Biochemistry-US 1990, 29, 10219.
%
\bibitem[11]{van_gunsteren}
Tironi, I.; Sperb, R.; Smith, P. E.; van Gunsteren, W. F.
J Chem Phys 1995, 102, 5451.
%
\bibitem[12]{zauhar}     
Zauhar, R. J.; Morgan, R. S. 
J Mol Biol 1985, 186, 815.                     
%
\bibitem[13]{juffer}
Juffer, A. H.;  Botta, E. F. F.; van Keulen, B. A. M.; van der Ploeg, A.;
Berendsen, H. J. C.
J Comput Phys 1991, 97, 144.
%
\bibitem[14]{still}
Still, W. C.; Tempczyk, A.; Hawley, R. C.;  Hendrickson, T.
J Am Chem Soc 1990, 112, 6127.  
%
\bibitem[15]{case}
Onufriev, A.; Bashford, D.; Case, D. A.
J Phys Chem B 2000, 104, 3712.
%
\bibitem[16]{curutchet}
Curutchet, C.; Orozco, M.; Luque, J. F.; Mennucci, B.; Tomasi, J.
J Comput Chem 2006, 27, 1769.                       
%
\bibitem[17]{jorgensen}
Carlson, H. A.; Jorgensen, W. L.       
J Phys Chem 1995, 99, 10667.
%
\bibitem[18]{connolly}
Connolly, M. L.                    
J Am Chem Soc 1985, 107, 1118. \\
http://www.netsci.org/Science/Compchem/feature14e.html
%
\bibitem[19]{lii}
Lii, J.-H.; Allinger, N. L.        
J Am Chem Soc 1989, 111, 8576.
%
\bibitem[20]{aqvist}
Aqvist, J.                   
J Phys Chem 1990, 94, 8021.
%
\bibitem[21]{weiner}
Weiner, S. J.; Kollman, P. A.; Case, D. A.; Singh, U. C.; Ghio, C.;
Alagona, G.; Profeta, S.; Weiner, P.
J Am Chem Soc 1984, 196, 765. 
%
\bibitem[22]{halgren}
Halgren, T.                
J Am Chem Soc 1992, 114, 7827.
%
\bibitem[23]{kollman}
Kollman, P. A.; Massova, I.; Reyes, C.; Kuhn, B.; Huo, S.; Chong, L.;
Lee, M.; Lee, T.; Duan, Y.; Wang, W.; Donini, Q.; Cieplak, P.; Srinivasan,
J.; Case, D. A.; Cheatham III, T. E.
Acc Chem Res 2000, 33, 889.
%
\bibitem[24]{bates}
Page, C. S.; Bates, P. A.
J Comput Chem 2006, 27, 1990.
%
\bibitem[25]{becke}
Becke, A. D.
J Chem Phys 1997, 107, 8554.
%
\bibitem[26]{sadlej}
Sadlej, A. J.
Theor Chim Acta 1991, 79, 123. 
%
\bibitem[27]{pdb}
Berman, H. M.; Westbrook, J.; Feng, Z.; Gilliland, G.; Bhat, T. N.; 
Weissig, H.; Shindyalov, I. N.; Bourne; P. E.
Nucleic Acids Res 2000, 28, 235.
%
\bibitem[28]{pdb1}
Akiyama, Y.; Onizuka, K.; Noguchi, T.; Ando, M.
Proc. 9th Genome Informatics Workshop (GIW'98), Universal Academy Press, 
1998, ISBN 4-946443-52-5, 131-140 \\
{\em http://mbs.cbrc.jp/pdbreprdb-cgi/reprdb\_menu.pl}
%
\bibitem[29]{vorobjev}
Vorobjev, Y. N.; Hermans, J.
Biophys J 1997, 73, 722.
%
\bibitem[30]{sanner}
Sanner, M. F.;  Olson, A. J.; Spehner, J.-C.
Proc. 11th ACM Symp Comp Geom, 1995, C6-C7.
%
\bibitem[31]{skrivanek}
Hayryan, S.; Hu, C.-K.; Sk\v{r}iv\'{a}nek, J.; Hayryan, E.; Pokorn\'{y}, I.
J Comput Chem 2005, 26, 334.   
%
\bibitem[32]{schaftenaar}
Schaftenaar, G.; Noordik, J. H.   
J Comput Aid Mol Des 2000, 14, 123.
%
\bibitem[33]{cornell}
Cornell, W. D.; Cieplak, P.; Bayly, C. I.; Gould, I. R.; Merz, K. M.;
Ferguson, D. M.; Spellmeyer, D.; Fox, T.; Caldwell, J. W.; Kollman, P. A.
J Am Chem Soc 1995, 117, 5179.
%
\bibitem[34]{hoefi}
H\"ofinger, S.
Modelling Molecular Structure and Reactivity in Biological Systems, 
RSC Publishing, 2006, ISBN 0-85404-668-2, 151.
%
\bibitem[35]{hoefi1}
H\"ofinger, S.
J Comput Chem 2005, 26, 1148.  
%
\bibitem[36]{marshall}
Marshall, N. J.; Grail, B. M.
J Pept Sci, 2000, 6, 186.
%
\bibitem[37]{tinker}
Ren, P.;  Ponder, J. W.
J Phys Chem B 2003, 107, 5933.
%
\bibitem[38]{gauss03}
Gaussian 03, Revision B.05,  
Frisch, M. J.; Trucks, G. W.; Schlegel, H. B.; Scuseria, G. E.; Robb, M. A.;
Cheeseman, J. R.; Montgomery Jr., J. A.; Vreven, T.; Kudin, K. N.; 
Burant, J. C.; Millam, J. M.; Iyengar, S. S.; Tomasi, J.; Barone, V.; 
Mennucci, B.; Cossi, M.; Scalmani, G.; Rega, N.; Petersson, G. A.; 
Nakatsuji, H.; Hada, M.; Ehara, M.; Toyota, K.; Fukuda, R.; Hasegawa, J.;
Ishida, M.; Nakajima, T.; Honda, Y.; Kitao, O.; Nakai, H.; Klene, M.; Li, X.;
Knox, J.E.; Hratchian, H. P.; Cross, J. B.; Bakken, V.; Adamo, C.; 
Jaramillo, J.; Gomperts, R.; Stratmann, R. E.; Yazyev, O.; Austin, A. J.;
Cammi, R.; Pomelli, C.; Ochterski, J. W.; Ayala, P. Y.; Morokuma, K.; 
Voth, G. A.; Salvador, P.; Dannenberg, J. J.; Zakrzewski, V. G.; Dapprich, S.;
Daniels, A. D.; Strain, M. C.; Farkas, O.; Malick, D. K.; Rabuck, A. D.;
Raghavachari, K.; Foresman, J. B.; Ortiz, J. V.; Cui, Q.; Baboul, A. G.;
Clifford, S.; Cioslowski, J.; Stefanov, B. B.; Liu, G.; Liashenko, A.;
Piskorz, P.; Komaromi, I.; Martin, R. L.; Fox, D. J.; Keith, T.; 
Al-Laham, M. A.; Peng, C. Y.; Nanayakkara, A.; Challacombe, M.; 
Gill, P. M. W.; Johnson, B.; Chen, W.; Wong, M. W.; Gonzalez, C.; Pople, J. A.
Gaussian, Inc., Wallingford CT, 2004.
%
\bibitem[39]{esqc-00}
European Summerschool in Quantum Chemistry, Book II, 
Bj\"orn O. Roos and Per-Olof Widmark, Eds, 2000, 561.             
%
\end{thebibliography}
\end{document}